\documentclass[sigconf]{acmart}

\AtBeginDocument{%
  \providecommand\BibTeX{{%
    \normalfont B\kern-0.5em{\scshape i\kern-0.25em b}\kern-0.8em\TeX}}}

\setcopyright{acmlicensed}
\copyrightyear{2023}
\acmYear{2023}
\acmDOI{XXXXXXX.XXXXXXX}

\acmConference[MSR 2024]{}{June 15--16,
  2024}{Lisbon, Portugal}
\acmISBN{978-1-4503-XXXX-X/18/06}

\usepackage{subcaption}
\usepackage{float}
\usepackage{xcolor}

\definecolor{boxdarkgray}{RGB}{66,66,66}
\definecolor{boxlightgray}{RGB}{238,238,238}

\makeatletter
\newcommand\footnoteref[1]{\protected@xdef\@thefnmark{\ref{#1}}\@footnotemark}
\makeatother

\newboolean{showcomments}
\setboolean{showcomments}{true}
\ifthenelse{\boolean{showcomments}}
  {\newcommand{\nb}[2]{
    \fbox{\bfseries\sffamily\scriptsize#1}
    {\sf\small$\blacktriangleright$\textit{#2}$\blacktriangleleft$}
   }
  }
  {\newcommand{\nb}[2]{}
  }

\newlength\MAX  

\usepackage{float}
\usepackage{floatpag}
\begin{document}

\title{Fixing Smart Contract Vulnerabilities: A Comparative Analysis of Literature and Developer's Practices}

\author{Francesco Salzano}
\orcid{0000-0002-1029-4861}
 \authornotemark[1]
\affiliation{%
  \institution{STAKE Lab, University of Molise}
  \city{Pesche}
   \country{Italy}
 }
 \email{f.salzano1@studenti.unimol.it}

\author{Simone Scalabrino}
\orcid{0000-0003-1764-9685}
\affiliation{%
 \institution{STAKE Lab, University of Molise}
 \city{Pesche}
 \country{Italy}
}
\email{simone.scalabrino@unimol.it}

\author{Rocco Oliveto}
\orcid{0000-0002-7995-8582}
\affiliation{%
  \institution{STAKE Lab, University of Molise}
 \city{Pesche}
 \country{Italy}
}
 \email{rocco.oliveto@unimol.it}

\author{Remo Pareschi}
\orcid{0000-0002-4912-582X}
\affiliation{%
  \institution{STAKE Lab, University of Molise}
 \city{Pesche}
 \country{Italy}
}
 \email{remo.pareschi@unimol.it}

\renewcommand{\shortauthors}{Salzano et al.}

\begin{abstract}
Smart Contracts are programs running logic in the Blockchain network by executing operations through immutable transactions. The Blockchain network validates such transactions, storing them into sequential blocks of which integrity is ensured. Smart Contracts deal with value stakes, if a damaging transaction is validated, it may never be reverted, leading to unrecoverable losses. To prevent this, security aspects have been explored in several fields, with research providing catalogs of security defects, secure code recommendations, and possible solutions to fix vulnerabilities. In our study, we refer to vulnerability fixing in the ways found in the literature as guidelines.  However, it is not clear to what extent developers adhere to these guidelines, nor whether there are other viable common solutions and what they are.
The goal of our research is to fill knowledge gaps related to developers' observance of existing guidelines and to propose new and viable solutions to security vulnerabilities. 
To reach our goal, we will obtain from Solidity GitHub repositories the commits that fix vulnerabilities included in the DASP TOP 10 and we will conduct a manual analysis of fixing approaches employed by developers.
Our analysis aims to determine the extent to which literature-based fixing strategies are followed. Additionally, we will identify and discuss emerging fixing techniques not currently documented in the literature. Through qualitative analysis, we will evaluate the suitability of these new fixing solutions and discriminate between valid approaches and potential mistakes.
\end{abstract}

\ccsdesc[500]{Software and its engineering}
\ccsdesc[500]{Security and privacy~Software security engineering}

\keywords{Smart Contract engineering, Smart Contract vulnerabilities, Smart Contract vulnerability fixing}


\maketitle

\section{Introduction}
Blockchain technology has become one of the most hyped technologies, gaining growing interest since the presentation of Bitcoin \cite{nakamoto2008bitcoin}. Thereafter, Smart Contracts (SCs), which are programs running logic on Blockchains, have increasingly gained adoption, becoming responsible for managing high stakes \cite{zou2019smart}. Vulnerabilities in the context of blockchain, are flaws or weaknesses in the design, implementation, or use of blockchain technologies that can be exploited to perform malicious or unwanted actions, as well as vulnerabilities existing in the code of SCs.
Such vulnerabilities can lead to significant losses in value, as in the case of the DAO attack, which led to the malicious withdrawal of cryptocurrencies worth about \$60 million \cite{porru2017blockchain}. 

As a consequence, security plays an essential role. Starting from this premise, several vulnerability detection tools have been developed and are available in the literature \cite{feist2019slither, tikhomirov2018smartcheck, ferreira2020smartbugs}, as well as empirical studies on their effectiveness \cite{durieux2020empirical, ghaleb2020effective}. These studies focused on Ethereum, in which Solidity serves as a predominant language \cite{rameder2022review}, and we follow the same setting in our work.

Research provided security smells, which are indicators that could indicate security problems that may have undesirable effects on the execution of SCs \cite{8859521}, as well as security defects, defined as errors that induce incorrect results \cite{chen2020defining},
moreover, security code recommendations refer to best practices and guidelines aimed at ensuring the security of software code that are available in the literature \cite{10123449}. 
Nonetheless, it is not clear to what extent the supplied fixing guidelines are followed nor if there are other valid fixing strategies used by developers.

In this work, we propose a study to fill this gap. SCs are still in their infancy, so security guidelines should be reviewed periodically. Developers may provide new solutions that can enrich existing approaches, which will be analyzed to state which are valid and which are erroneous. First, we collect vulnerability fix recommendations from the literature. Second, we plan to mine Solidity GitHub repositories searching for commits that fix vulnerabilities checking if these fixes are included in the suggestions available in the literature to understand the extent to which they adhere to them. 

As a second analysis, we collect vulnerability fixes different from those that are already known in the literature. In this analysis, we critically analyze whether found solutions are suitable for the community.
To validate our proposal, we executed a simplified version of the designed experimental plan (with a single evaluator and limiting the mining process to the first 1000 repositories). 

In summary, our study will provide insights into vulnerability correction strategies used by SC developers and introduce new valid correction approaches.  

\section{Background and Related Work}
\label{sec:background}

\textbf{Blockchain.}
The Blockchain is a self-governed peer-to-peer network transaction system that allows for secure execution of operations eliminating the need for a trusted third party \cite{8716424}. Transactions are executed on a decentralized ledger composed of sequential blocks linked by themselves, with an immutable connection to the predecessor, ensuring the chain integrity.
Each block stores validated transactions according to consensus algorithms.
The ledger is shared and replicated, and participants of the network can read and write data on it, granting transparent access to its stored data for every network participant.
\\ 
\textbf{Smart Contracts.}
The SC concept was introduced in the 1990s by Szabo, initially described as computerized protocols that executed in transactions the terms of a contract  \cite{szabo1997formalizing}. Contemporary interpretations consider SCs as event-driven software replicated over decentralized nodes in equal copies, which are set to automatically execute code when certain conditions are met \cite{zou2019smart}. As well as Blockchains, SCs are immutable. Although SCs can be made updatable through the use of a proxy that routes calls to a new implementation, the original contract remains published on the blockchain, maintaining its immutability \cite{bodell2023proxy}.

Users or other SCs can interact with SCs by calling them via transactions. Nodes in the Blockchain network validate the transactions, when a transaction is valid the result of the execution of the logic codified in the SC is written on their local copy of the Blockchain. To reach inclusion in a block, all the nodes must execute this logic in the same way, now stored data are irreversible due to the immutability of the Blockchain. This implies that if a transaction finishes unexpectedly, the result may be not revertible.
\\
\textbf{Ethereum.}
Ethereum is the second largest Blockchain, after Bitcoin, but the second most adopted as a SC environment, enabling SC execution through the Ethereum Virtual Machine, which compiles into Ethereum bytecode the SCs written in high-level languages, of which Solidity is the predominant one \cite{zou2019smart, buterin2014next}.
\\
\textbf{Ethereum's Gas.}
Gas in Ethereum is the unit used to measure the work done by Ethereum to carry out interaction within the network. SCs are run by miners on their nodes, in return they receive a quantity of gas as a reward. This reward is paid by users who request transactions; each transaction has a \textit{gas limit} that determines the maximum gas cost. Whether such a cost overcomes the limit, the transaction will be reverted, also raising an exception \cite{chen2020defining}.
\\
\textbf{Smart Contract Vulnerabilities.}
Research refers to the Decentralized Application Security Project (DASP)\footnote{https://dasp.co/} TOP 10 SC vulnerabilities when classifying security vulnerabilities \cite{durieux2020empirical, ferreira2020smartbugs, dia2021empirical}. Vulnerabilities included in the DASP are listed in Table~\ref{tab:vulnerabilities} along with a description.

\begin{table*}[h]
        \centering
         \caption{DASP Vulnerabilities with description.}
        \begin{tabular}{p{0.15\textwidth}p{0.8\textwidth}}
            \toprule
            \textbf{Category} & \textbf{Description} \\
            \midrule
            
            Reentrancy & Contracts can call other contracts. Reentrancy is caused when the target contract is recursively called by an external contract before finishing to update its state after the initial execution, thus leveraging an inconsistent state.  \\
            
            Access Control & 
            The lack of secure access and authorization to functions gives attackers direct ways to access private values or functions.\\
            
            Arithmetic & Math operations are performed on variables with fixed dimensions. Numbers that exceed these dimensions overflow or underflow. When exploited, Arithmetic vulnerability gives incorrect results, compromising reliability.\\
            
            Unchecked Calls & Solidity offers low-level calls such as \textit{call()}. The error of these calls is not propagated, not reverting the current execution. Instead, they return a Boolean value set to false, which if it is not checked, can lead to undesirable outcomes. \\
            Denial of Service & 
            Several ways could lead to denials of service, such as maliciously increasing the gas required to compute a function, for instance, sending an array with a huge dimension to a function that loops over it. In this case, if gas block limitations are exceeded, transactions will be reverted. \\
            
            Bad Randomness &
            Randomness is hard to get right in Blockchains due to the need for consensus. The sources of randomness inner to Solidity are predictable. Thus, malicious users can exploit relying on such predictability.
            \\
            
            Front Running & 
            Transactions require a period before they are added to a block. An attacker can view the transaction pool and may include another transaction block before the original one. This mechanism can be exploited to reorder transactions to the attacker's advantage.\\   
            
            Time Manipulation & Decision are often brought according to time-related conditions. The current time is usually obtained using \textit{block.timestamp} or \textit{now} instructions. But this value comes from the miners and can be manipulated maliciously by them. \\  
            
            Short Address & 
            Solidity pads shorter arguments to 32 bytes, thus, an attacker may manipulate the data sent, making the SC read more data than was sent.
            \\  
            
            Unknowns &
            The DASP TOP 10 also indicates a category of vulnerabilities not yet known. \\     
            \bottomrule
        \end{tabular}
       
        \label{tab:vulnerabilities}

\end{table*}

\textbf{Related Work.} SCs need a high level of security assessment before being deployed. Zou et al. conducted a survey, finding that most of the respondents to this survey declared that SC development has a fairly higher requirement for code security compared to traditional development due to the management of digital assets and the irreversible nature of the transactions \cite{zou2019smart}.
Given these motivations, SC security has played a crucial part in academic research. Thus besides static analyzers \cite{feist2019slither, tikhomirov2018smartcheck, ferreira2020smartbugs}, fuzz tools have fed the suite of SC detector tools \cite{9000089} as well Machine Learning and Deep Learning-based tools \cite{shakya2022smartmixmodel, zhang2022reentrancy}.

Duriex et al. conducted an empirical review evaluating vulnerability detectors \cite{durieux2020empirical}, in the same context, Ghaleb and Pattabiraman evaluated static analysis tools by injecting security-related bugs \cite{ghaleb2020effective}. both suggest the need to improve their detection performance.
Nonetheless, despite the wide set of vulnerability detectors, developers still depend on finding vulnerabilities manually \cite{ghaleb2022towards}.

Therefore, providing developers with security smells and vulnerability mitigation guidelines was crucial. Indeed, several researchers have contributed. Demir et al. reviewed the literature searching for vulnerabilities that must be sidestepped, also creating a catalog of security smells \cite{8859521}. Chen et al. have gone deeper, not only providing an extended set of smells, encompassing those that are concerned with security; but also providing solutions for each vulnerability  \cite{chen2020defining}.

Furthermore, Nguyen et al. and Chen et al. presented \textit{SGUARD} and \textit{TIPS}, respectively, two automated approaches to patch SCs; in their research, they also provided some fixing patterns \cite{nguyen2021sguard, chen2023tips}.
The most recent related work is conducted by Zhou et al., who developed \textit{SmartREP}, a one-line fixing technique for SC repair \cite{10123449}. Alongside, they provided 13 code recommendations to fix vulnerabilities resulting from the literature review and commonly encountered in SC development, paving the way for studies of bug-related code changes explicitly.

Based on the knowledge gathered from the above research, we will refer to them to identify known security vulnerability-fixing approaches, moreover, we provide an online appendix with a catalog of literature guidelines \footnote{https://github.com/francescosalzano1/literature-fixing-guidelines-for-Smart-Contracts/tree/main}. From this point on, we will refer to these approaches as guidelines in the literature. We will consider as a guideline only those fixing approaches that provide the safe code. During this stage, we built up the guideline set on the knowledge obtained by papers resulting from keyword-based research, searching for "smart contract security fix". We considered sources from IEEE, ACM, ScienceDirect, and SpringerLink digital libraries. In addition, we only considered peer-reviewed journal and conference papers written in English, and we excluded studies where Solidity was not the language used in the SCs. The guidelines are derived from these research considering the code provided within, thus we extracted the resolution patterns by thoroughly analyzing these resolution models and categorizing them according to the DASP TOP 10 taxonomy.

\section{Empirical Study Design}
The \textit{goal} of the study we propose is to evaluate whether developers fix SC security vulnerabilities following the current research guidelines and whether there are valid fixes not supplied by the literature.
The \textit{perspective} is of researchers interested in SC security. The \textit{context} will consist of a dataset of fixing security vulnerability commits in public Solidity SC repositories.

In detail, to reach our goal, the proposed study is guided by the following research questions:
\label{sec:study}
\begin{itemize}	
    \item \emph{RQ{1}: To what extent do developers adhere to the fixing guidelines in the literature? }
    \item \emph{RQ{2}: Which are the valid fixing approaches besides those found in the literature?}

\end{itemize}
\subsection{Study Context}
\label{sec:study_context}

The context of our research will be represented by a dataset composed of commits that resolve security vulnerabilities included in the DASP TOP 10. Fixing commits will provide pairs of vulnerable and fixed code, supplying insightful fixing procedures. These commits will be selected from all Solidity repositories that will pass our next-defined filters and a further phase of manual analysis. This dataset will be created by mining commits from GitHub since the birth of Ethereum. Detailed procedures to select the sample of interest and the processes devoted to selecting vulnerability-fixing commits in the remainder part of this section.


\subsubsection{Finding Candidate Repositories}

We will focus our study on Solidity SCs. Therefore, we will start performing Data Collection by getting repositories from GitHub, and applying filters to repositories while collecting them. Indeed, we will consider solely repositories that present content in Solidity language and with a star count greater or equal to 10. This choice aims to significantly reduce the number of repositories by selecting the ones most appreciated by the community, to have a limited number of repositories with good quality, following a common practice \cite{dabic2021sampling, rosa2018quality}. 
In addition, we motivate the choice with several factors, indeed, star count is a good proxy for assessing the quality of a repository \cite{dabic2021sampling}. Moreover, This approach allows us to focus on a manageable number of repositories that are more likely to be both relevant and of higher quality.

Given that the SCs are independent, we will not set a lower bound about the number of files contained in the repositories, although we do not expect repositories with a single file to respond to the star-based filter. 

The number of repositories returned using GitHub's API for searching repositories with the given filters on language and number of stars is currently 5874, so given the handleable total, we will consider all of them.

\subsubsection{Mining Commits}

Thereafter, starting from the gathered repositories we will mine commits using \textit{PyDriller}, a framework that helps developers to extract information from Git software repositories \cite{spadini2018pydriller}. Solidity files are characterized by the extension \textit{.sol}, so, we will only store commits that have modified at least one file with this extension to conduct the next steps. Moreover, exclusively commits having commit messages dealing with vulnerability fixes will be included in the study, following the filtering procedure that we will explain further.
Additionally, we will exclude \textit{merge commits} and duplicates.

\subsubsection{Filtering Commits}

Mined commits will be filtered using an \textit{NLP-based filter}. The filter will be implemented leveraging \textit{SpaCy},\footnote{https://spacy.io/} an open-source NLP library, by employing \textit{SpaCy} lemmatization to get the lemmas of the commit messages. In detail, the filter will be applied to each commit message, which will be tokenized and cleaned of stopwords. We will implement the NLP-based filter in such a way as to accept commit messages containing the lemma of vulnerability and the lemma "fix", for each vulnerability exposed in~\ref{tab:vulnerabilities}, for instance, commit messages like "\textit{Arithmetic vulnerability fixed}". 
As a result, this approach will provide a restricted dataset with an adequate number of filtered commits to execute further procedures.

An additional filter will be applied, in detail, we will exclude files that will not contain the \textit{pragma solidity} declaration which indicates to the Solidity compiler which version of the Solidity language to use to compile the contract, because it could indicate that these contracts are intended for import purposes only.
At the end of this process, collected commits will be considered candidate commits, the resulting dataset of developers' changes counts 3462 instances of modified Solidity SCs to be further manually analyzed.

\subsection{Experimental Procedure}
\label{sec:evaluation_procedure}
In this section, we detail the experimental procedure that will follow to answer our \emph{RQs}. 

\begin{figure}[!htbp]
    \centering
    \includegraphics[width=\linewidth]
        {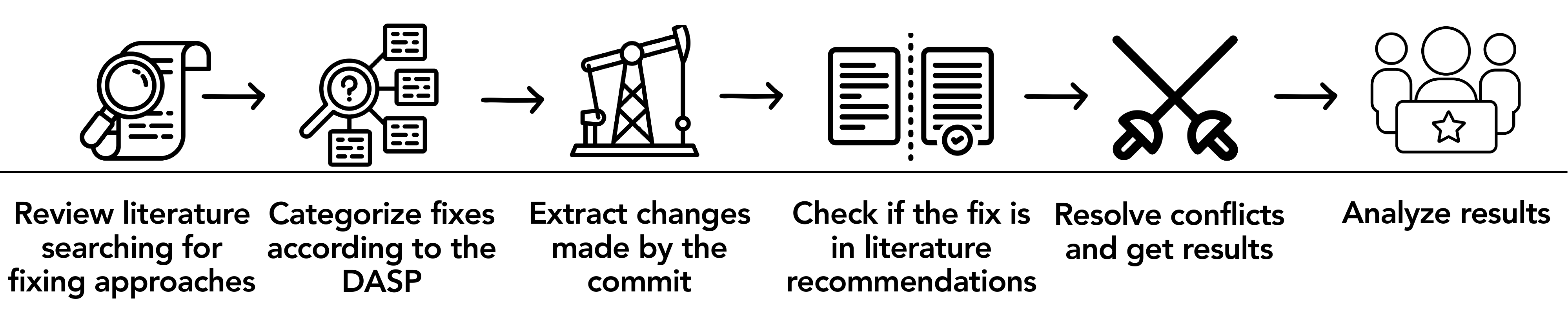}
    \caption{Overall workflow to answer RQ1}
    \label{fig:rq1}
\end{figure}

Fig~\ref{fig:rq1} depicts the overall workflow that we will carry on to address \emph{RQ1}.

To answer the question \emph{RQ1}, we will start by having two of the authors manually analyze the candidate commits.
Commits involving more than 3 files will be excluded if the commit message does not specify the vulnerable file, aiming to have guidance from the developers who patched the contracts, make our analysis more scalable, and reduce the subjective practices that will be conducted by validators.
Moreover, commits with messages that will not specify what type of vulnerability is fixed will be excluded unless, by examining the commit manually, the analyzers will be able to assign the vulnerability type with certainty.

Hence, the two analyzers will label the commits separately, where the two labels will be equal, the commit will be tagged with the common value; elsewhere conflict will be resolved. In detail, whether the two analyzers label a commit with different vulnerability types, it will be necessary to decide which label is more appropriate. This choice will be taken following a discussion between the two evaluators. The labels used will match the DASP TOP 10 labels.
At the end of this phase, the remaining commits will be selected for experimentation.
Then, for each commit in our manually analyzed dataset, we will extract the SC before and after the fixing commit. At this point, for each commit, we have the pre-fix and post-fix versions. 

Code changes will be collected by comparing the two versions, i.e., pre-commit and post-commit code, with the \textit{diff\_parsed} property of the dictionary returned by Pydriller, which represents a single commit. Such a dictionary has 2 keys: “added” and “deleted”, containing the added and deleted lines, respectively. Moreover, the same difference can be viewed on GitHub searching for the commit changes, helping us review the changes made through the graphical interface, which highlights modifications.

Then, once we obtain the difference, we will consider as a fix a change in which is modified at least one row of the SC that contained the vulnerability; differences related to spaces, indentation, and empty rows will be ignored. In addition, fixing implies changes, if the vulnerable lines are just removed, the changes will not be considered a fix. Whether the commits brought several changes, the evaluators will find the fix and isolate it from the rest of the made changes. 

Thereafter, the evaluators will ascertain whether the difference between pre-fix and post-fix Solidity SC can be attributed to the mitigations available in the literature.

To this purpose, the two evaluating authors will analyze the commit instances independently, stating separately whether each instance analyzed contains collected resolution models from previous research.
In case of conflict, the authors will discuss until they reach a consensus. 
Detailing this step, once each evaluator conducted an independent analysis of the commit instances if any discrepancies arose in their assessments, the evaluators documented the specific points of contention, clearly outlining their differing perspectives on whether a given change could be attributed to mitigations available in the literature. After contrasts discussion, whether the evaluators will not reach for a common choice, the instance will be analyzed by a third evaluator. The inter-rater reliability between the two evaluators will be reported using Cohen's kappa coefficient to measure the level of agreement \cite{cohen1960coefficient}.

\begin{figure}[!htbp]
    \centering
    \includegraphics[width=\linewidth]
        {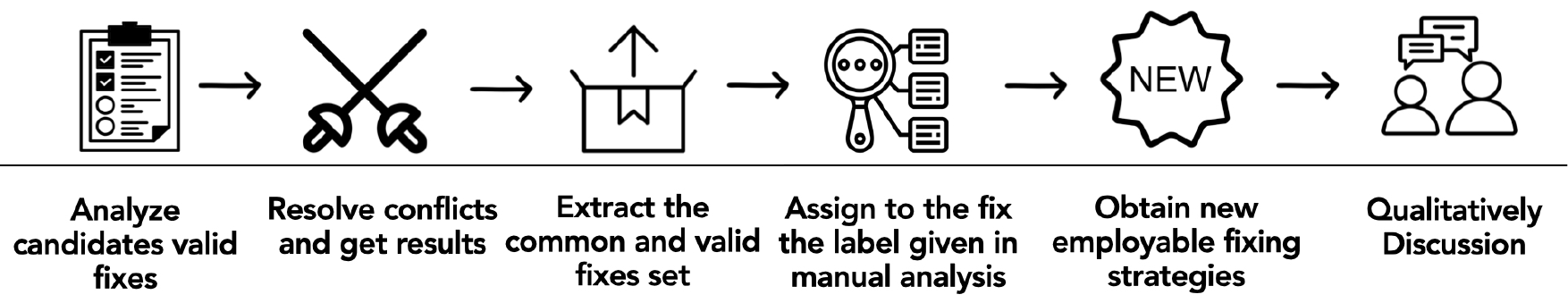}
    \caption{Overall workflow to answer RQ2}
    \label{fig:rq2}
\end{figure}
At the end of this step, we will provide results showing how many fixes are under literature recommendations. To address \emph{RQ1}, we will report the number and percentage of fixing commits that adopt approaches known in the literature, categorized by each category of DASP, and also provide the most fixed vulnerabilities. Then, for each category, the computed percentage will report the extent to which developers adhere to literature fixing guidelines. Fixing approaches not included in the collection of literature guidelines will serve for further analysis to answer \emph{RQ2}.

\begin{figure*}[!htbp]
    \centering
    \includegraphics[width=\linewidth]
        {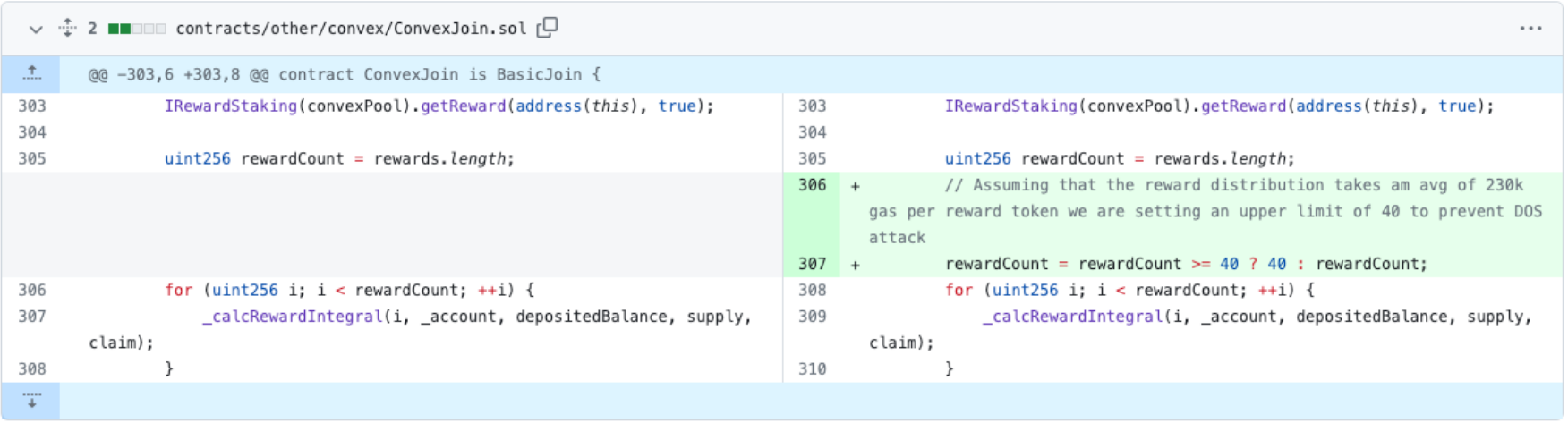}
    \caption{Denial of service patched by a commit.}
    \label{fig:commit}
    \vspace{-0.2cm}
\end{figure*}

The plan dedicated to responding to \emph{RQ2} involves analyzing changes made by developers that fix vulnerabilities among those that are not included in the set of literature recommendations. Hence the designed workflow to address \emph{RQ2} is shown in Fig ~\ref{fig:rq2}.

These changes will be critically analyzed by at least two SC practitioner evaluators; they will search for valid fixing approaches. 
We define a valid fixing approach as a solution developers use to address a security issue that has been evaluated and deemed appropriate for the given kind of vulnerability.
Even in this case, the authors who will evaluate must reach a consensus. 
Qualitative metrics involved in this analysis will include adaptability and applicability of the fix, analyzing its ability to adapt in ever-changing scenarios as the SC one and to be flexible in different contexts. Moreover, long-lasting implications will be considered, indeed, some corrections may solve the immediate problem but may be subject to future problems.

Emerged valid fixes will be categorized according to the label assigned to the commit by the evaluators in the manual analysis stage. We will provide both new employable fixing strategies and a detailed qualitative discussion of each common fix found. Additionally, we will describe the resolution approach and explain its validity. To answer \emph{RQ2} we will report each kind valid of patch that emerged along with the related discussion result.

\subsection{Preliminary Results}

In order to have a preliminary validation of our experimental procedure and to obtain preliminary results, we performed the experimental plan following the procedures outlined above. The entire plan was followed strictly, except for introducing a limit to the commit mining process and using only one evaluator. Specifically, only commits from the first 1000 repositories responding to the filters regarding the number of stars and content in Solidity were mined. 
We have decided to conduct
this initial experiment to obtain some preliminary results that could make our report more credible. Nonetheless,
to eliminate any potential bias introduced by this choice, we will carry out the final experiment by ignoring these
initial results and following a double-checked experimental plan. In this preliminary analysis, we preferred more repositories (1k) and fewer evaluators rather than having more evaluators and fewer instances, in a way that avoids effortful cross-checking.
Preliminary results appeared promising, Fig~\ref{fig:commit} shows an instance, in which developers used a fix that we do not observe in the reviewed literature. In detail, the function was vulnerable to \textit{denial of service} because the unbounded iteration can consume more gas than the block gas limit, permanently reverting the transaction. 
The change made by the commit limits the upper limit, and the gas consumption becomes bounded, resolving the \textit{denial of service}. Thus, we marked this fix as valid.

\section{Limitations, Challenges and Mitigation }

\textbf{Wrong Vulnerability Fixes.} Some of the changes made by developers may not be valid fixes. To mitigate this risk, we will first rely on line guidance from the literature, on organizations such as \textit{Consensys} \footnote{https://consensys.io/}, and on our experience, in addition, we will validate each fix by double-checking. As a final mitigation point, research evidence has shown that SC practitioners have more awareness of security than traditional software developers \cite{wan2021smart}, therefore, we can assume that most of the fixes will be valid.
\\
\textbf{Bias in Validating New Fixing Approaches.} To state if the changes made by a fixing commit could be challenging. Still, to mitigate this risk, we will double-check the emerged fixes after a critical analysis.

\label{sec:limitations}

\section{Conclusion}
\label{sec:conclusion}
In recent years, SCs have gained increasing adoption, and security vulnerabilities have been a central point in research. Vulnerability detectors are widespread \cite{di2023smartbugs, feist2019slither}, nonetheless, SC developers rely on identifying vulnerabilities by themselves, demonstrating an elevated awareness of security-related tasks. Previous research provided fixing guidelines \cite{chen2020defining, 10123449, nguyen2021sguard}, our study will help evaluate not solely whether developers follow these recommendations, but also in providing different and new options to fix vulnerabilities. Datasets and results that will emerge from our study will be publicly available to enable foster research, paving the way in expanding the possibilities of fixes by auto-repair tools and the deeper study of fixing changes. 

\begin{acks}
    This project was partially funded through TruMaN (Italian Ministry of University and Research, 2022, PRIN, Project 2022F5CLN2).
\end{acks}

\bibliographystyle{unsrt}
  \bibliography{main}

\begin{thebibliography}{10}

\bibitem{nakamoto2008bitcoin}
Satoshi Nakamoto.
\newblock Bitcoin: A peer-to-peer electronic cash system.
\newblock {\em Decentralized Business Review}, page 21260, 2008.

\bibitem{zou2019smart}
Weiqin Zou, David Lo, Pavneet~Singh Kochhar, Xuan-Bach~Dinh Le, Xin Xia, Yang Feng, Zhenyu Chen, and Baowen Xu.
\newblock Smart contract development: Challenges and opportunities.
\newblock {\em IEEE Transactions on Software Engineering}, 47(10):2084--2106, 2019.

\bibitem{porru2017blockchain}
Simone Porru, Andrea Pinna, Michele Marchesi, and Roberto Tonelli.
\newblock Blockchain-oriented software engineering: challenges and new directions.
\newblock In {\em 2017 IEEE/ACM 39th International Conference on Software Engineering Companion (ICSE-C)}, pages 169--171. IEEE, 2017.

\bibitem{feist2019slither}
Josselin Feist, Gustavo Grieco, and Alex Groce.
\newblock Slither: a static analysis framework for smart contracts.
\newblock In {\em 2019 IEEE/ACM 2nd International Workshop on Emerging Trends in Software Engineering for Blockchain (WETSEB)}, pages 8--15. IEEE, 2019.

\bibitem{tikhomirov2018smartcheck}
Sergei Tikhomirov, Ekaterina Voskresenskaya, Ivan Ivanitskiy, Ramil Takhaviev, Evgeny Marchenko, and Yaroslav Alexandrov.
\newblock Smartcheck: Static analysis of ethereum smart contracts.
\newblock In {\em Proceedings of the 1st international workshop on emerging trends in software engineering for blockchain}, pages 9--16, 2018.

\bibitem{ferreira2020smartbugs}
Jo{\~a}o~F Ferreira, Pedro Cruz, Thomas Durieux, and Rui Abreu.
\newblock Smartbugs: A framework to analyze solidity smart contracts.
\newblock In {\em Proceedings of the 35th IEEE/ACM International Conference on Automated Software Engineering}, pages 1349--1352, 2020.

\bibitem{durieux2020empirical}
Thomas Durieux, Jo{\~a}o~F Ferreira, Rui Abreu, and Pedro Cruz.
\newblock Empirical review of automated analysis tools on 47,587 ethereum smart contracts.
\newblock In {\em Proceedings of the ACM/IEEE 42nd International conference on software engineering}, pages 530--541, 2020.

\bibitem{ghaleb2020effective}
Asem Ghaleb and Karthik Pattabiraman.
\newblock How effective are smart contract analysis tools? evaluating smart contract static analysis tools using bug injection.
\newblock In {\em Proceedings of the 29th ACM SIGSOFT International Symposium on Software Testing and Analysis}, pages 415--427, 2020.

\bibitem{rameder2022review}
Heidelinde Rameder, Monika Di~Angelo, and Gernot Salzer.
\newblock Review of automated vulnerability analysis of smart contracts on ethereum.
\newblock {\em Frontiers in Blockchain}, 5:814977, 2022.

\bibitem{8859521}
Mehmet Demir, Manar Alalfi, Ozgur Turetken, and Alexander Ferworn.
\newblock Security smells in smart contracts.
\newblock In {\em 2019 IEEE 19th International Conference on Software Quality, Reliability and Security Companion (QRS-C)}, pages 442--449, 2019.

\bibitem{chen2020defining}
Jiachi Chen, Xin Xia, David Lo, John Grundy, Xiapu Luo, and Ting Chen.
\newblock Defining smart contract defects on ethereum.
\newblock {\em IEEE Transactions on Software Engineering}, 48(1):327--345, 2020.

\bibitem{10123449}
Xiaocong Zhou, Yingye Chen, Hanyang Guo, Xiangping Chen, and Yuan Huang.
\newblock Security code recommendations for smart contract.
\newblock In {\em 2023 IEEE International Conference on Software Analysis, Evolution and Reengineering (SANER)}, pages 190--200, 2023.

\bibitem{8716424}
Shikah~J. Alsunaidi and Fahd~A. Alhaidari.
\newblock A survey of consensus algorithms for blockchain technology.
\newblock In {\em 2019 International Conference on Computer and Information Sciences (ICCIS)}, pages 1--6, 2019.

\bibitem{szabo1997formalizing}
Nick Szabo.
\newblock Formalizing and securing relationships on public networks.
\newblock {\em First monday}, 1997.

\bibitem{bodell2023proxy}
William~E Bodell~III, Sajad Meisami, and Yue Duan.
\newblock Proxy hunting: Understanding and characterizing proxy-based upgradeable smart contracts in blockchains.
\newblock In {\em 32nd USENIX Security Symposium (USENIX Security 23)}, pages 1829--1846, 2023.

\bibitem{buterin2014next}
Vitalik Buterin et~al.
\newblock A next-generation smart contract and decentralized application platform.
\newblock {\em white paper}, 3(37):2--1, 2014.

\bibitem{dia2021empirical}
Bruno Dia, Naghmeh Ivaki, and Nuno Laranjeiro.
\newblock An empirical evaluation of the effectiveness of smart contract verification tools.
\newblock In {\em 2021 IEEE 26th Pacific Rim International Symposium on Dependable Computing (PRDC)}, pages 17--26. IEEE, 2021.

\bibitem{9000089}
Bo~Jiang, Ye~Liu, and W.K. Chan.
\newblock Contractfuzzer: Fuzzing smart contracts for vulnerability detection.
\newblock In {\em 2018 33rd IEEE/ACM International Conference on Automated Software Engineering (ASE)}, pages 259--269, 2018.

\bibitem{shakya2022smartmixmodel}
Supriya Shakya, Arnab Mukherjee, Raju Halder, Abyayananda Maiti, and Amrita Chaturvedi.
\newblock Smartmixmodel: machine learning-based vulnerability detection of solidity smart contracts.
\newblock In {\em 2022 IEEE international conference on blockchain (Blockchain)}, pages 37--44. IEEE, 2022.

\bibitem{zhang2022reentrancy}
Zhuo Zhang, Yan Lei, Meng Yan, Yue Yu, Jiachi Chen, Shangwen Wang, and Xiaoguang Mao.
\newblock Reentrancy vulnerability detection and localization: A deep learning based two-phase approach.
\newblock In {\em Proceedings of the 37th IEEE/ACM International Conference on Automated Software Engineering}, pages 1--13, 2022.

\bibitem{ghaleb2022towards}
Asem Ghaleb.
\newblock Towards effective static analysis approaches for security vulnerabilities in smart contracts.
\newblock In {\em 37th IEEE/ACM International Conference on Automated Software Engineering}, pages 1--5, 2022.

\bibitem{nguyen2021sguard}
Tai~D Nguyen, Long~H Pham, and Jun Sun.
\newblock Sguard: towards fixing vulnerable smart contracts automatically.
\newblock In {\em 2021 IEEE Symposium on Security and Privacy (SP)}, pages 1215--1229. IEEE, 2021.

\bibitem{chen2023tips}
Qianguo Chen, Teng Zhou, Kui Liu, Li~Li, Chunpeng Ge, Zhe Liu, Jacques Klein, and Tegawend{\'e}~F Bissyand{\'e}.
\newblock Tips: towards automating patch suggestion for vulnerable smart contracts.
\newblock {\em Automated Software Engineering}, 30(2):31, 2023.

\bibitem{dabic2021sampling}
Ozren Dabic, Emad Aghajani, and Gabriele Bavota.
\newblock Sampling projects in github for msr studies.
\newblock In {\em 2021 IEEE/ACM 18th International Conference on Mining Software Repositories (MSR)}, pages 560--564. IEEE, 2021.

\bibitem{rosa2018quality}
Giovanni Rosa, Simone Scalabrino, Gabriele Bavota, and Rocco Oliveto.
\newblock What quality aspects influence the adoption of docker images?
\newblock {\em ACM Transactions on Software Engineering and Methodology}, 2018.

\bibitem{spadini2018pydriller}
Davide Spadini, Maur{\'\i}cio Aniche, and Alberto Bacchelli.
\newblock Pydriller: Python framework for mining software repositories.
\newblock In {\em Proceedings of the 2018 26th ACM Joint meeting on european software engineering conference and symposium on the foundations of software engineering}, pages 908--911, 2018.

\bibitem{cohen1960coefficient}
Jacob Cohen.
\newblock A coefficient of agreement for nominal scales.
\newblock {\em Educational and psychological measurement}, 20(1):37--46, 1960.

\bibitem{wan2021smart}
Zhiyuan Wan, Xin Xia, David Lo, Jiachi Chen, Xiapu Luo, and Xiaohu Yang.
\newblock Smart contract security: A practitioners' perspective.
\newblock In {\em 2021 IEEE/ACM 43rd International Conference on Software Engineering (ICSE)}, pages 1410--1422. IEEE, 2021.

\bibitem{di2023smartbugs}
Monika di~Angelo, Thomas Durieux, Jo{\~a}o~F Ferreira, and Gernot Salzer.
\newblock Smartbugs 2.0: An execution framework for weakness detection in ethereum smart contracts.
\newblock {\em arXiv preprint arXiv:2306.05057}, 2023.

\end{thebibliography}
\end{document}